# Modeling and Simulation of 2D Transducers Based on Suspended Graphene-Based Heterostructures in Nanoelectromechanical Pressure Sensors

Quan Liu[1,2,3], Chang He[1,3], Jie Ding[4]*, Wendong Zhang[5,6]*, and Xuge Fan[1,3,4]*

[1] Advanced Research Institute for Multidisciplinary Science, Beijing Institute of Technology, Beijing 100081, China.

[2] Yangtze Delta Region Academy of Beijing Institute of Technology, Jiaxing 314003, China.

[3] Center for Interdisciplinary Science of Optical Quantum and NEMS Integration, Beijing Institute of Technology, 100081 Beijing, China.

[4] School of Integrated Circuits and Electronics, Beijing Institute of Technology, Beijing 100081, China.

[5] State Key Laboratory of Dynamic Measurement Technology, North University of China, Taiyuan 030051, China.

[6] National Key Laboratory for Electronic Measurement Technology, School of Instrument and Electronics, North University of China, Taiyuan 030051, China.

*E-mail: xgfan@bit.edu.cn, jie.ding@bit.edu.cn, wdzhang@nuc.edu.cn

**ABSTRACT:** Graphene-based 2D heterostructures exhibit excellent mechanical and electrical properties, which are expected to exhibit better performances than graphene for nanoelectromechanical pressure sensors. Here, we built the pressure sensor models based on suspended heterostructures of graphene/h-BN, graphene/$MoS_2$, and graphene/$MoSe_2$ by using COMSOL Multiphysics finite element software. We found that suspended circular 2D membranes show the best sensitivity to pressures compared to rectangular and square ones. We simulated the deflections, strains, resonant frequencies, and Young's moduli of suspended graphene-based heterostructures under the conditions of different applied pressures and geometrical sizes, built-in tensions, and the number of atomic layers of 2D membranes. The Young's moduli of 2D heterostructures of graphene, graphene/h-BN, graphene/$MoS_2$, and graphene/$MoSe_2$ were estimated to be 1.001TPa, 921.08 GPa, 551.11 GPa, and 475.68 GPa, respectively. We also discuss the effect of highly asymmetric cavities on device performance. These results would contribute to the understanding of the mechanical properties of graphene-based heterostructures and would be helpful for the design and manufacture of high-performance NEMS pressure sensors.

**INTRODUCTION**

Graphene has an ultra-thin thickness (0.335 nm for single-layer graphene),[1] Young's modulus of up to 1 Tpa,[2] fracture strength of up to 130 GPa[3], and an electron mobility of up to $2.5 \times 10^5$ $cm^2$ /V s,[4] and exhibits ultra-strong van der Waals attraction between graphene and $SiO_2$ surface.[5] Such ultra-thin thickness, excellent mechanical and electrical properties of graphene make it a promising material for ultra-small and sensitive NEMS pressure sensors. [6,7]

In 2013, A. D. Smith et al.[8] fabricated an ultra-small and sensitive NEMS pressure sensor based on a monolayer-suspended graphene membrane.[9] In 2015, an ultra-sensitive and small graphene squeeze-film pressure sensor was reported.[10] In 2016, the highly sensitive capacitive pressure sensor based on ultra-large suspended graphene membrane was realized.[11] A static capacitive pressure sensor using a graphene drum was reported in 2017, with the detection of capacitance changes down to 50 aF and pressure differences down to 25 mbar.[12] An ultrasensitive resonant pressure sensor based on a sealed graphene nanodrum was reported in 2019, with an average responsivity of up to 39.2 kHz/mbar.[13] In 2020, sensitive capacitive pressure sensors based on suspended graphene membrane arrays were fabricated with a sensitivity of up to 47.8 aF Pa$^{-1}$ mm$^{-2}$.[14] In 2021, the pirani pressure sensor based on suspended graphene membranes with small footprint was realized.[15] Silicon nitride membranes or polymers were used to assist suspended graphene membranes for improving device yields of pressure sensors.[16–20]

Beyond graphene, other 2D materials also show promising properties for NEMS applications, such as their relatively high in-plane stiffness and strength.[21] For instance, Young's modulus of monolayer h-BN, MoS$_2$ and MoSe$_2$ are reported to be 865 GPa, 270 GPa and 177 GPa, respectively.[21] The piezoresistive gauge factors of MoS$_2$ and MoSe$_2$ have been reported to be about -148 ± 19[22] and 1800[23] respectively, two to three orders of magnitude higher than commonly reported values in graphene.[8,16,24] Therefore, MoS$_2$ and MoSe$_2$ can be used to build heterostructures of graphene/MoS$_2$ and graphene/MoSe$_2$ as transducers for NEMS pressure sensors to potentially improve the sensitivity. h-BN can be used to build graphene/h-BN heterostructures to potentially improve the stability and yield of graphene-based pressure sensors. However, applications of 2D materials beyond graphene in NEMS have been rarely reported.[25–30]

In this work, we built the models of pressure sensors based on graphene and graphene-based heterostructures and studied the deformation behaviors, deflections, strains, resonant frequencies and Young's moduli of graphene and graphene-based heterostructures under different conditions by simulating them with COMSOL 6.1 finite element simulation software.

**RESULTS AND DISCUSSION**

In this paper, we designed the schematic diagram of the NEMS pressure sensors, which are composed of the silicon substrate, etched silicon dioxide layer with a cavity, suspended heterostructures of graphene/h-BN, graphene/$MoS_2$, graphene/$MoSe_2$, and electrodes (**Figure 1**a). The partially magnified structures and the cross-section of the designed NEMS pressure sensors were shown in **Figure 1**b and c, respectively. We modeled the section of the suspended 2D heterostructures over the cavity of the $SiO_2$ layer in COMSOL (**Figure 1**d). And we simulated the deflections, strains, resonant frequencies and Young's moduli of suspended graphene-based heterostructures under the conditions of different shapes of 2D membranes, geometrical sizes, built-in tensions, atomic layers and applied pressures.

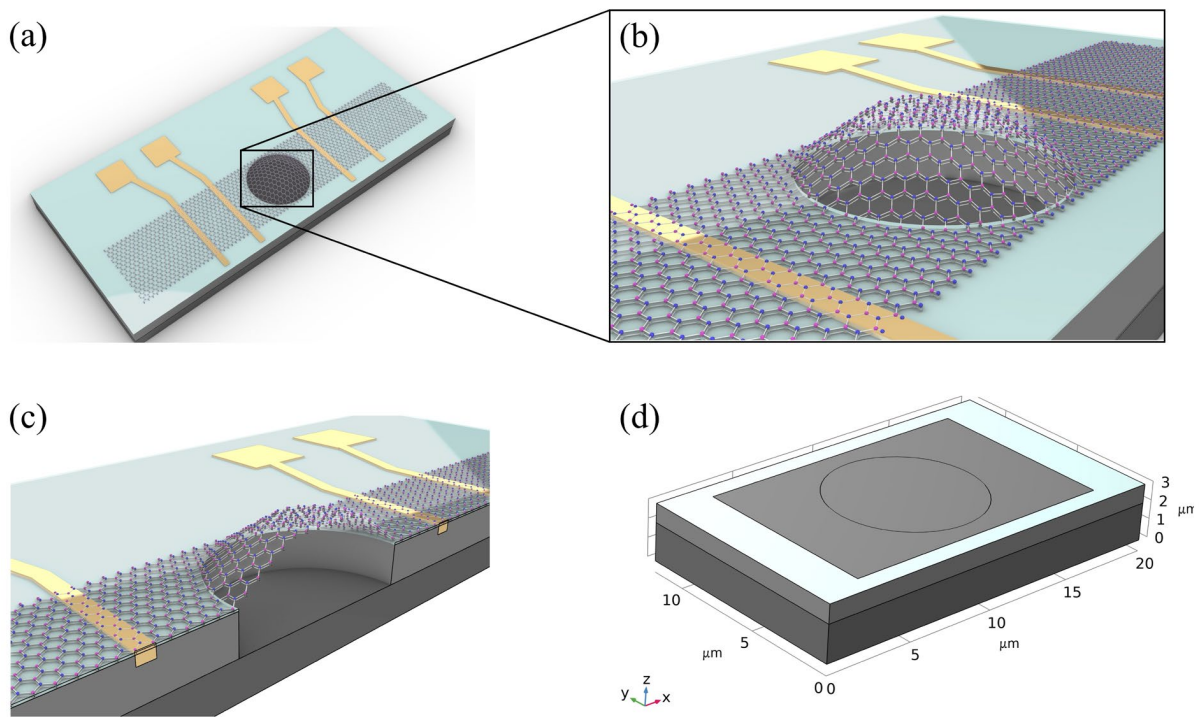

**Figure 1**. Schematics of NEMS pressure sensors based on graphene-based heterostructures. (a) Schematic diagram of the designed pressure sensors based on suspended graphene-based heterostructures. (b) Partial magnification of the designed pressure sensors. (c) Cross-section

view of the designed pressure sensors. (d) The Simplified model of the designed pressure sensors for COMSOL simulation.

The model consists of the shell interface in the structural mechanics module in COMSOL, which can be used to analyze the mechanical properties of suspended graphene-based heterostructures. Geometrical nonlinearities of suspended graphene-based heterostructure should be considered in the study since the deflection of graphene-based heterostructure is much larger than its membranes' thickness. The material parameters for simulations were illustrated in **Table 1**.

Table 1. Material property settings.[21]

|  | Young's moduli (TPa) | Poisson's ratio | Density (kg/m$^3$) | Membrane thickness (nm) |
|---|---|---|---|---|
| Graphene | 1 | 0.16 | 2200 | 0.335 |
| MoS$_2$ | 0.270 | 0.27 | 5060 | 0.65 |
| MoSe$_2$ | 0.177 | 0.23 | 6900 | 0.65 |
| h-BN | 0.865 | 0.225 | 2290 | 0.325 |

**Shape of 2D Membranes**

We modeled circular, square, and rectangular membranes of graphene-based heterostructures that were suspended over cavities (**Figure 2**a). Graphene/h-BN heterostructures were chosen as the representative and simulated to study the impact of the shape of cavities on the deflection and strain by applying a series of pressures. The maximum deflection occurs in the center of the membrane, indicated by the deepest color on the color bar (**Figure 2**a). The maximum strain of the membrane occurs in the center of the membrane.

The area of different shapes of 2D membranes of graphene/h-BN heterostructures is set to be 16π μm², and the built-in tension is set to be 0.1N/m. This is because the average built-in tension of a fully-clamped suspended graphene membrane on a circular cavity was measured to be 0.085 N/m based on nanoindentation measurements.[31] Additionally, the average built-in stress of a doubly-clamped suspended graphene ribbon was measured to be about 300 MPa.[32–35] As shown in **Figure 2**b and c, the circular membrane of the graphene/h-BN heterostructure had the largest maximum deflection and strain compared to the square and rectangular ones. And the maximum deflection and strain of the square membrane of the graphene/h-BN heterostructure are larger than rectangular one. For instance, at the applied pressure of 0.1 MPa, the maximum deflections of circular, square, and rectangular membranes of graphene/h-BN heterostructures were 0.23 μm, 0.21654 μm, and 0.13946 μm (**Figure** 2b), with the corresponding maximum strains of 0.406%, 0.362%, and 0.206%, respectively (**Figure 2**c). Therefore, it can be estimated that the circular membranes of 2D heterostructures has the best responsivity to the pressure. Therefore, the circular membranes of graphene-based heterostructures will be chosen for the related simulations.

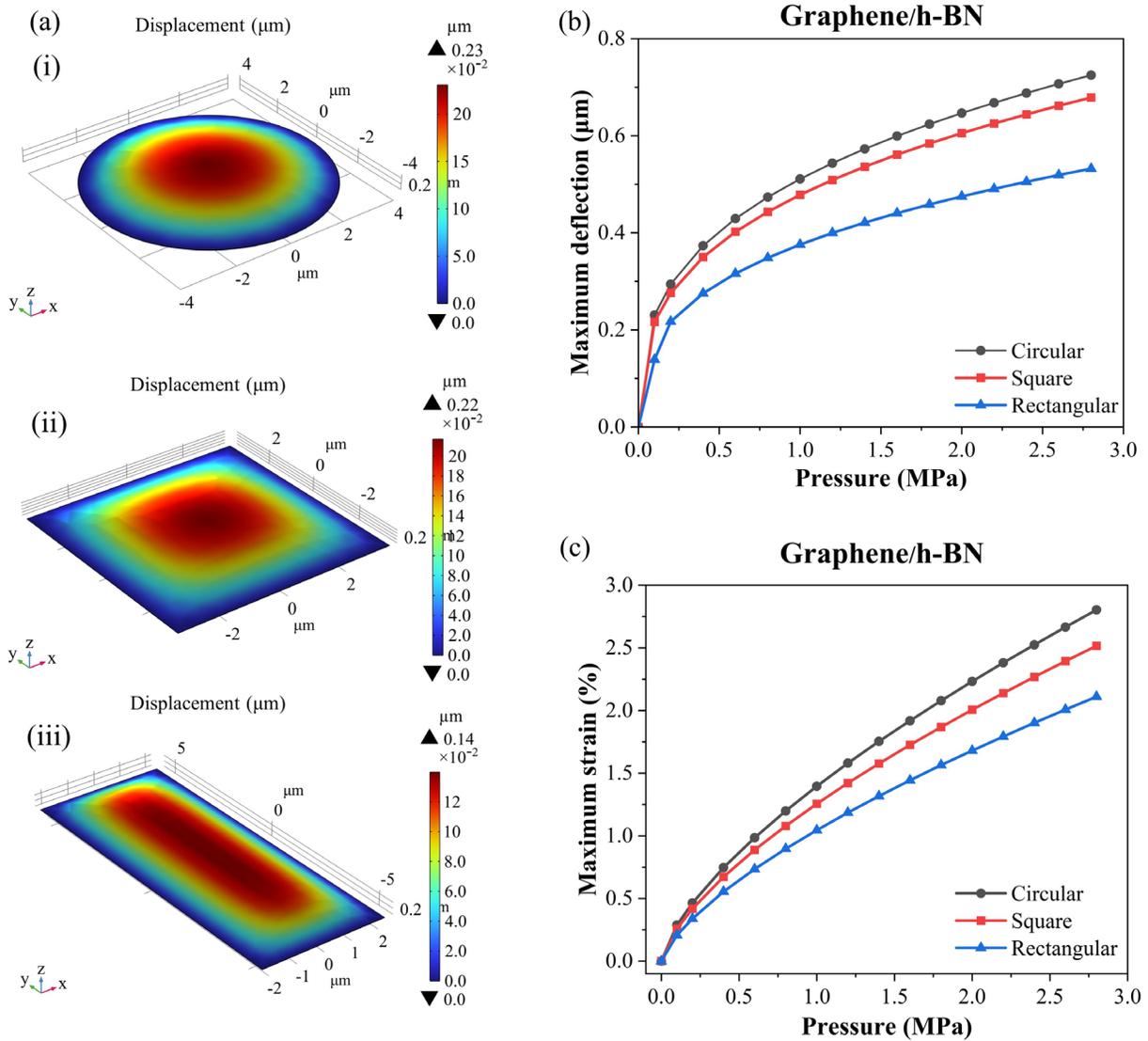

**Figure 2.** Deflections and strains of graphene/h-BN heterostructures with different shapes. (a) Modeling of 2D membranes of suspended graphene/h-BN heterostructures with the same area but different membrane's shapes: (i) Circular membrane, (ii) Square membrane, (iii) Rectangular membrane. And the simulation of the maximum deflections (b) and the maximum strains (c) of circular, square, and rectangular membranes of graphene/h-BN heterostructures with the same area of 16π μm² versus applied pressures under the condition of the built-in tension of 0.1N/m.

**Deflections of 2D Membranes**

For the suspended circular 2D membranes, the maximum membrane deflection due to the applied pressure can be described by[36]

$$P = \frac{4t}{a^2}\sigma_0\delta + \frac{2.67t}{a^4}\frac{E}{1-v}\delta^3 \quad (1)$$

where $P$ is the pressure, $\delta$, $a$, $t$, $E$, $v$, $\sigma_0$ is the maximum deflection, radius, thickness, Young's modulus, Poisson's ratio, and built-in tension of the circular 2D membranes, respectively.

We used graphene and heterostructures of graphene/h-BN, graphene/MoS$_2$, and graphene/MoSe$_2$ as sensitive membranes of pressure sensors to simulate the membranes' maximum deflections under different conditions (**Figure 3**). **Figure 3**a illustrates the maximum deflections of monolayer graphene and 2D heterostructures of monolayer graphene with monolayer h-BN, MoS$_2$ and MoSe$_2$ obviously increased with increasing the applied pressures under the conditions of the built-in tension of 0.1 N/m and the 2D membrane's radius of 4 μm. At the same applied pressures, the maximum deflection of the monolayer graphene membranes is larger than those of all heterostructures. Further, the maximum deflection of the heterostructure of monolayer graphene with monolayer MoSe$_2$ is larger than the heterostructure of monolayer graphene with monolayer MoS$_2$. And the heterostructure of monolayer graphene with monolayer h-BN has the smallest deflection. For instance, the maximum deflections of monolayer graphene membrane, heterostructures of monolayer graphene with monolayer h-BN, MoS$_2$ and MoSe$_2$ were 0.28444 μm, 0.23035 μm, 0.24461 μm, and 0.25687 μm at the applied pressure of 0.1MPa. In addition, as illustrated in **Figure 3**a, the relationship between the maximum deflection and the applied pressure was approximately linear as the applied pressures ranged from 0 to 0.1 MPa. This is because the cubic term $\frac{2.67t}{a^4}\frac{E}{1-v}\delta^3$ in equation (1) can be ignored as the applied pressures were small. As the applied pressure was larger than 0.1 MPa, the cubic term $\frac{2.67t}{a^4}\frac{E}{1-v}\delta^3$ in equation (1) will be larger than the primary term $\frac{4t}{a^2}\sigma_0\delta$, resulting in a nonlinear relationship between the maximum deflection and the applied pressure.

To study the impact of geometrical sizes and built-in tensions on the deflections of 2D membranes, we performed corresponding simulations. The maximum deflections of monolayer graphene and 2D heterostructures of monolayer graphene with monolayer h-BN, MoS$_2$ and MoSe$_2$ visibly increased with increasing the radiuses of 2D membranes under the conditions of the applied pressure of 0.1 MPa and built-in tension of 0.1 N/m (**Figure 3**b), but slowly decreased with increasing the built-in tensions of 2D membranes under the conditions of the applied pressure of 0.1 MPa and membrane's radius of 4 μm (**Figure 3**c).

To study the impact of the thickness of 2D membranes on the deflection, we simulated the maximum deflection of graphene membranes with different atomic layers and 2D heterostructures of monolayer graphene with different atomic layers of h-BN, MoS$_2$, and MoSe$_2$ under the conditions of the applied pressure of 0.1 Mpa, built-in tension of 0.1 N/m, and the 2D membranes' radius of 4 μm. **Figure 3**d shows that the maximum deflections of all 2D membranes visibly deceased with increasing their thickness. Further, the maximum deflection of 2D heterostructure of monolayer graphene with the certain atomic layers of MoSe$_2$ is larger than both graphene membranes and 2D heterostructures of monolayer graphene with MoS$_2$ and h-BN that have the same number of atomic layers.

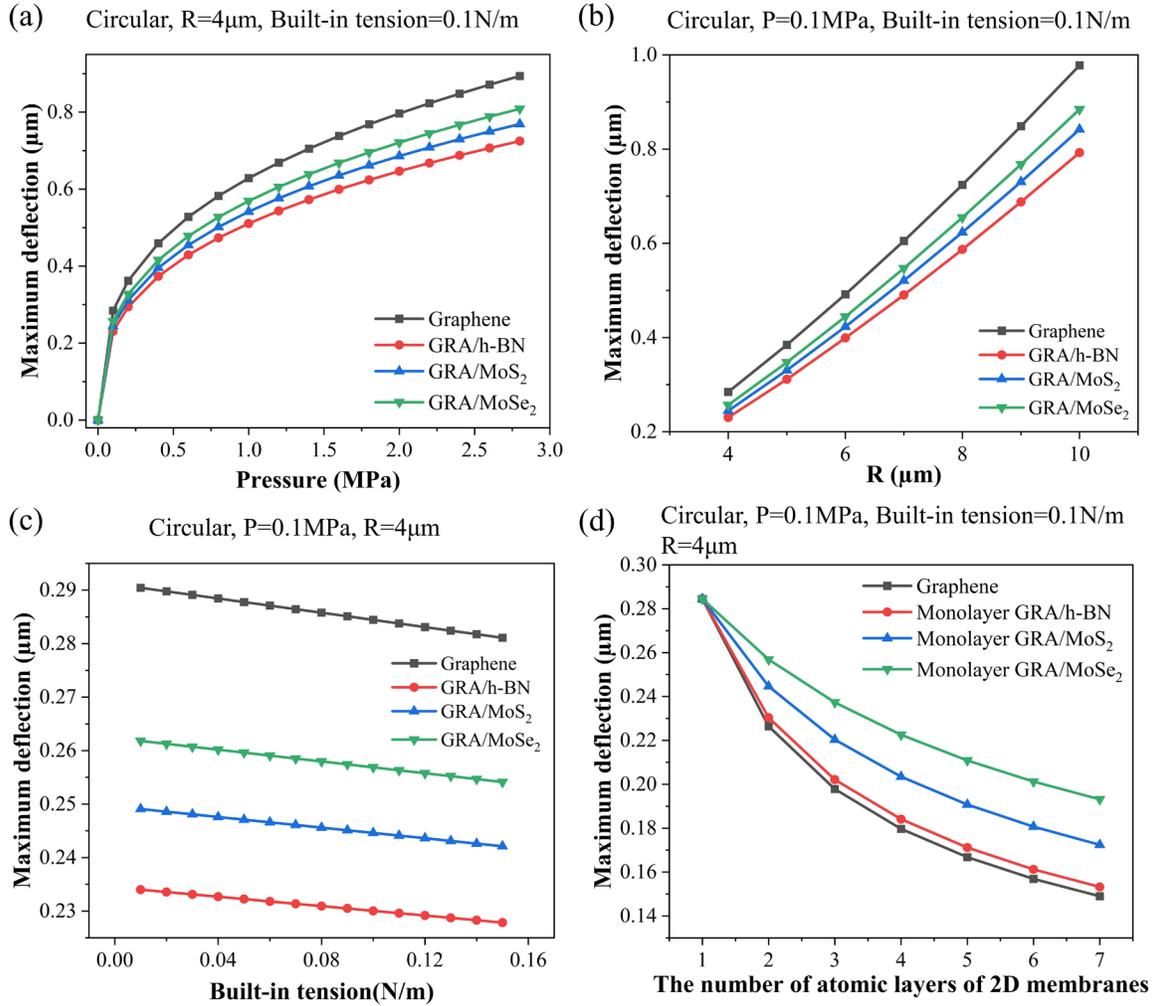

**Figure 3**. Maximum deflections of circular 2D membranes of graphene and graphene-based heterostructures under different conditions. (a) The maximum deflections of monolayer graphene and heterostructures of monolayer graphene with monolayer h-BN, $MoS_2$, and $MoSe_2$ versus applied pressures under the conditions of the built-in tension of 0.1 N/m and the 2D membrane's radius of 4 μm. (b) The maximum deflections of monolayer graphene and 2D heterostructures of monolayer graphene with monolayer h-BN, $MoS_2$ and $MoSe_2$ versus the radiuses of 2D membranes under the conditions of the applied pressure of 0.1 MPa and the built-in tension of 0.1 N/m. (c) The maximum deflections of monolayer graphene and 2D heterostructures of monolayer graphene with monolayer h-BN, $MoS_2$ and $MoSe_2$ versus built-in tensions under the conditions of the applied pressure of 0.1 MPa and 2D membranes' radius of 4 μm. (d) The maximum deflections of graphene membranes with different atomic layers and 2D heterostructures of monolayer graphene with different atomic-layers of h-BN, MoS$_2$ and MoSe$_2$ under the conditions of the applied pressure of 0.1 MPa, built-in tension of 0.1 N/m, and 2D membranes' radius of 4 μm. It should be noted that as the number of atomic layers of 2D membranes is 1, only monolayer graphene membrane is present without heterostructures.

**Strains of 2D Membranes**

The strain is an important parameter for analyzing the change of the resistance of the graphene and graphene-based 2D heterostructures, and the relationship between the resistance change and the strain can be described as[24]

$$GF = \frac{\Delta R/R}{\varepsilon} \tag{2}$$

where $\varepsilon$ is the strain, *GF* is the piezoresistive gauge factor, *R* is the resistance, and *ΔR* is the resistance change.

For the suspended membranes of graphene-based 2D heterostructures, the change of 2D membrane's strain due to the applied pressure can be described by[36]

$$P = \frac{4t\sigma_0}{a}\sqrt{\frac{3\varepsilon}{2}} + \frac{4t}{a}\frac{E}{1-v}\sqrt{\frac{3\varepsilon^3}{2}} \tag{3}$$

where *P* is the pressure, $\varepsilon$ is the strain, *a, t, E, v,* $\sigma_0$ is the radius, thickness, Young's modulus, Poisson's ratio, and built-in stress of circular 2D membranes, respectively.

We simulated the maximum strains of graphene membranes and graphene-based 2D heterostructures under different conditions. A strain profile of the suspended 2D heterostructure of monolayer graphene with monolayer h-BN induced by pressures was illustrated in **Figure 4**a.

The maximum strains of monolayer graphene and 2D heterostructures of monolayer graphene with monolayer h-BN, MoS$_2$ and MoSe$_2$ obviously increased with increasing the applied pressures under the conditions of the built-in tension of 0.1 N/m and the 2D membrane's radius of 4 μm (**Figure 4**b). At the same applied pressures, the maximum strain of the monolayer graphene membrane is larger than those of all heterostructures. Further, the maximum strain of the heterostructure of monolayer graphene with monolayer MoSe$_2$ is larger than the heterostructure of monolayer graphene with monolayer MoS$_2$. And the heterostructure of monolayer graphene with monolayer h-BN has the smallest membrane's strain. For instance, the maximum strain of monolayer graphene membranes and heterostructures of monolayer graphene with monolayer h-BN, MoS$_2$ and MoSe$_2$ are 0.335%, 0.23%, 0.2587% and 0.2862% at the applied pressure of 0.1MPa, respectively.

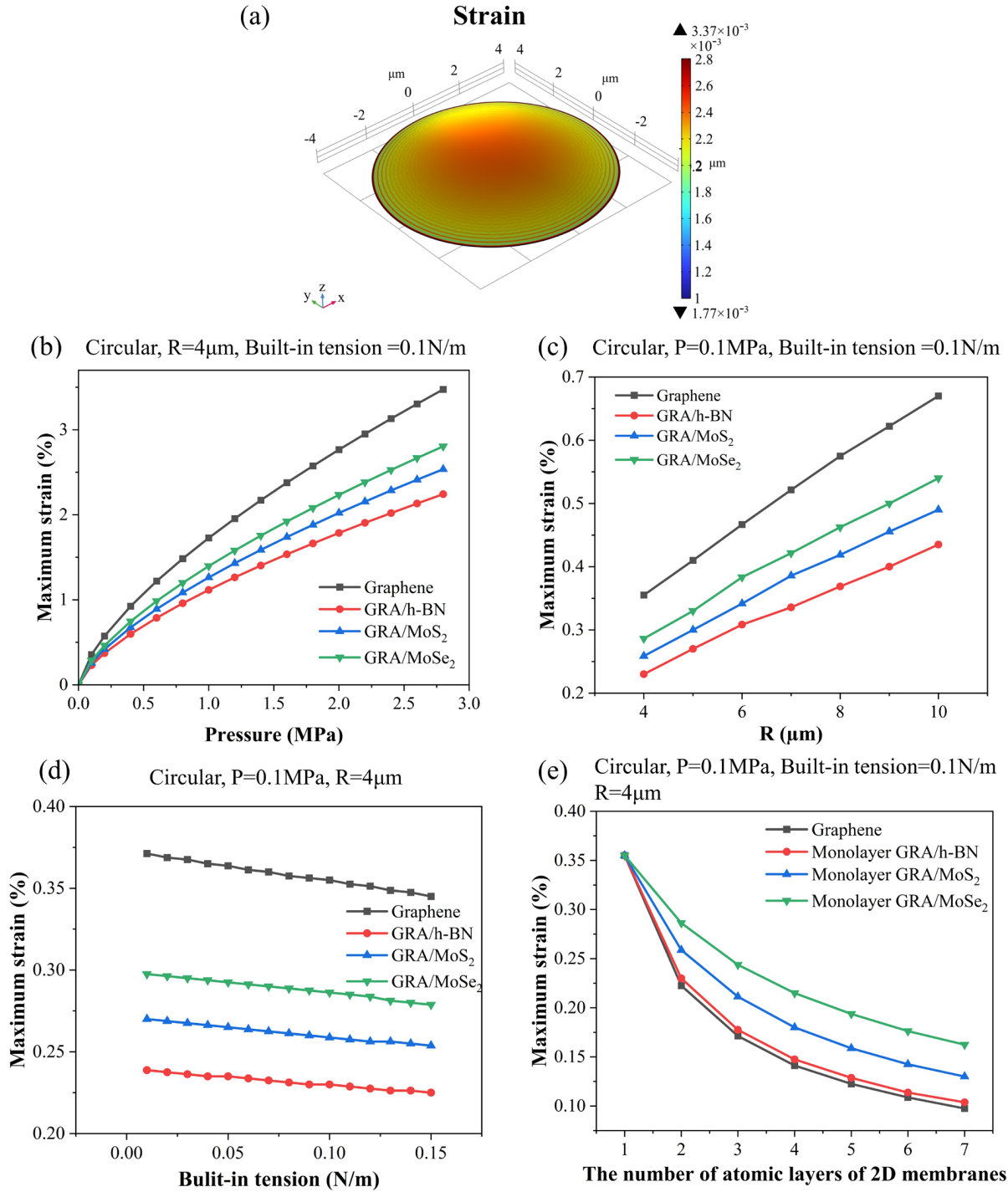

**Figure 4**. Strains of circular 2D membranes of graphene and graphene-based heterostructures under different conditions. (a) The strain profiles of the suspended heterostructure of monolayer graphene with monolayer h-BN at different applied pressures. (b) The maximum strains of monolayer graphene and 2D heterostructures of monolayer graphene with monolayer h-BN, MoS$_2$ and MoSe$_2$ versus applied pressures under the conditions of the built-in tension of 0.1 N/m and the 2D membrane's radius of 4 μm. (c) The maximum strains of monolayer graphene and 2D heterostructures of monolayer graphene with monolayer h-BN, MoS$_2$ and MoSe$_2$ versus the radiuses of 2D membranes under the conditions of the applied pressure of 0.1 Mpa and the built-in tension of 0.1 N/m. (d) The maximum strains of monolayer graphene and 2D heterostructures of monolayer graphene with monolayer h-BN, MoS$_2$ and MoSe$_2$ versus built-in tensions under the conditions of the applied pressure of 0.1 Mpa and the 2D membranes' radius of 4 μm. (e) The maximum strains of graphene membranes with different atomic layers and 2D heterostructures of monolayer graphene with different atomic layers of h-BN, MoS$_2$ and MoSe$_2$ under the conditions of the applied pressure of 0.1 Mpa, built-in tension of 0.1 N/m, and the 2D membranes' radius of 4 μm. It should be noted that as the number of atomic layers of 2D membranes is 1, only monolayer graphene membrane is present without heterostructures.

Furthermore, as illustrated in **Figure 4**b, the relationship between the heterostructure of monolayer graphene with monolayer h-BN has the smallest membrane's maximum strain and the pressure is close to linearity, compared to the relationship between the maximum deflection and the pressure, which would be useful for the practical applications of piezoresistive NEMS pressure sensors based on 2D membranes. This is because the relationship between the maximum strain and the pressure is mainly determined by the term $\frac{4t}{a}\frac{E}{1-v}\sqrt{\frac{3\varepsilon^3}{2}}$ of equation (3), which can be approximated as a kind of linear relationship in strain analysis. The relationship between maximum strain and resistance change can therefore be described as

$$\Delta R/R = GF \times k \times P \qquad (4)$$

where *k* is the ratio factor between the strain and the pressure.

To study the impact of geometrical sizes and built-in tensions on the maximum strains, we performed corresponding simulations. The maximum strains of monolayer graphene and 2D heterostructures of monolayer graphene with monolayer h-BN, MoS$_2$ and MoSe$_2$ fast increase with increasing the radiuses of 2D membranes under the conditions of the applied pressure of 0.1 MPa and built-in tension of 0.1 N/m (**Figure 4**c), but slowly increase with decreasing the built-in tensions of 2D membranes under the conditions of the applied pressure of 0.1 MPa and membrane's radius of 4 μm (**Figure 4**d).

To study the impact of the thickness of 2D membranes on the maximum strain, we simulated the maximum strain of graphene membranes with different atomic layers and 2D heterostructures of monolayer graphene with different atomic layers of h-BN, MoS$_2$ and MoSe$_2$, under the conditions of the applied pressure of 0.1 MPa, built-in tension of 0.1 N/m and the 2D membranes' radius of 4 μm. As shown in **Figure 4**e, the maximum strains of all 2D membranes visibly decreased with increasing their thicknesses. Further, the maximum strain of 2D heterostructure of monolayer graphene with the certain atomic layers of MoSe$_2$ is larger than both graphene membranes and 2D heterostructures of monolayer graphene with MoS$_2$ and h-BN that have the same number of atomic layers.

**Resonant Frequencies of 2D Membranes**

As the pressure is applied on the suspended graphene-based heterostructures, their resonant frequency are changed. The fundamental resonant frequency of the circular 2D membranes is given by the following equation[37]

$$f_0 = \frac{2.404}{2\pi a} \sqrt{\frac{S_p + S_0}{\rho}} \qquad (5)$$

where $f_0$ is the fundamental resonant frequency, $S_0$ is the initial tension of the membrane, and $S_p$ is the tension induced by the applied pressure, $a$ is the radius of the membrane, $\rho$ is is the density of membranes.

We simulated the resonant frequencies of membranes of graphene and graphene-based heterostructures under different conditions. The resonant modes at the first, second, third, fourth, fifth and sixth orders of the suspended heterostructure of monolayer graphene with monolayer h-BN were shown in **Figure 5**a, with corresponding resonant frequencies of 168.63 MHz, 195.39 MHz, 195.46 MHz, 241.6 MHz, 241.69 MHz and 265.81 MHz.

The resonant frequencies of monolayer graphene and 2D heterostructures of monolayer graphene with monolayer h-BN, MoS$_2$ and MoSe$_2$ obviously increased with increasing the applied pressures under the conditions of the built-in tension of 0.1 N/m and the 2D membrane's radius of 4 μm (**Figure 5**b). This indicated that the suspended membranes of graphene and graphene-based heterostructures can be used as resonant transducers for high-performance resonant NEMS pressure sensors. At the same applied pressures, the resonant frequency of the monolayer graphene membrane is larger than those of all heterostructures. Further, the resonant frequency of the heterostructure of monolayer graphene with monolayer h-BN is larger than the heterostructure of monolayer graphene with monolayer MoS$_2$. And the heterostructure of monolayer graphene with monolayer MoSe$_2$ has the smallest resonant frequency. For instance, the resonant frequencies of monolayer graphene membranes and heterostructures of monolayer graphene with monolayer h-BN, MoS$_2$ and MoSe$_2$ are 214.56 MHz, 168.63 MHz, 99.25 MHz and 84.99 MHz at the applied pressure of 0.1MPa.

Furthermore, the impact of the built-in tension on the resonant frequency is negligible when $S_p$ is larger than $S_0$ (equation (5)). Therefore, as $S_p$ is larger than $S_0$, the resonant frequency can be approximately expressed by the following equation[37]

$$f_0^3 \approx \frac{2.404^3}{64\pi^3} P \sqrt{\frac{8Et}{3\rho^3 a^4 (1-v)}} \tag{6}$$

where, $f_0$ is the fundamental resonant frequency, P is the pressure, $a$, $\rho$, $E$, $t$ is the radius, density, Young's modulus, and thickness of the suspended circular 2D membranes, respectively.

To study the impact of geometrical sizes and built-in tensions on the resonant frequencies, we performed corresponding simulations. The resonant frequencies of monolayer graphene and 2D heterostructures of monolayer graphene with monolayer h-BN, $MoS_2$ and $MoSe_2$ obviously decrease with increasing the radiuses of 2D membranes under the conditions of the applied pressure of 0.1 MPa and built-in tension of 0.1 N/m (**Figure 5**c), but slightly decreased with increasing the built-in tensions of 2D membranes under the conditions of the applied pressure of 0.1 MPa and membrane's radius of 4 μm (**Figure 5**d).

To study the impact of the thickness of 2D membranes on the resonant frequency, we simulated the resonant frequency of graphene membranes with different atomic layers and heterostructures of monolayer graphene with different atomic layers of h-BN, $MoS_2$ and $MoSe_2$, under the conditions of the applied pressure of 0.1 MPa, built-in tension of 0.1 N/m, and the 2D membranes' radius of 4 μm. The resonant frequencies of all 2D membranes visibly decreased with increasing their thicknesses (**Figure 5**e). The resonant frequency of the heterostructure of monolayer graphene with the certain atomic layers of h-BN is almost identical to that of graphene membranes with the same atomic layers, both of which are much larger than those of heterostructures of monolayer graphene with $MoS_2$ and $MoSe_2$ that have the same atomic layers. Further, the resonant frequency of the heterostructure of monolayer graphene with the certain atomic layers of MoS₂ is larger than that of the heterostructure of monolayer graphene with the same atomic layers of MoSe₂.

It can be seen from equation (5) that the resonant frequency increases with increasing the built-in tension. However, as the built-in tension $S_0$ is gradually increased and large enough, the tension $S_p$ induced by the applied pressure will be decreased, which would probably result in the decrease of the resonant frequency. This deduction has been verified by our simulation (**Figure 5**f). In addition, we also consider the possible inhomogeneity of the built-in tension in two-dimensional materials during the preparation process. For instance, to simulate the relationship of resonant frequency versus applied pressure, we set the built-in tensions of the left side and right side of the suspended circular graphene membrane to be 0.03 N/m and 0.06 N/m, as well as 0.1 N/m and 0.06 N/m, respectively (**Figure 5**f). The simulation results indicate that although the absolute values of the resonant frequencies under the conditions of inhomogeneous built-in tensions above are different with those under the conditions of the homogeneous built-in tensions, their overall trends of the relationship of resonant frequency versus applied pressure are similar to those under the conditions of the homogeneous built-in tension (**Figure 5**f).

As the applied pressure is small, the resonant frequency of the graphene/h-BN heterostructure increases with increasing built-in tension, and can be expressed by the following equation (7).[37]

$$f_0 = \frac{2.404t}{2\pi a} \sqrt{\frac{E}{\rho(1-v)\varepsilon} + \frac{t\sigma_0}{\rho}}, S_P \approx 0 \qquad (7)$$

where, $f_0$ is the fundamental resonant frequency, P is the pressure, $\varepsilon$, $a$, $\rho$, $E$, $t$, $v$, $\sigma_0$ is the strain, radius, density, Young's modulus, thickness, Poisson's ratio and built-in stress of circular 2D membranes, respectively.

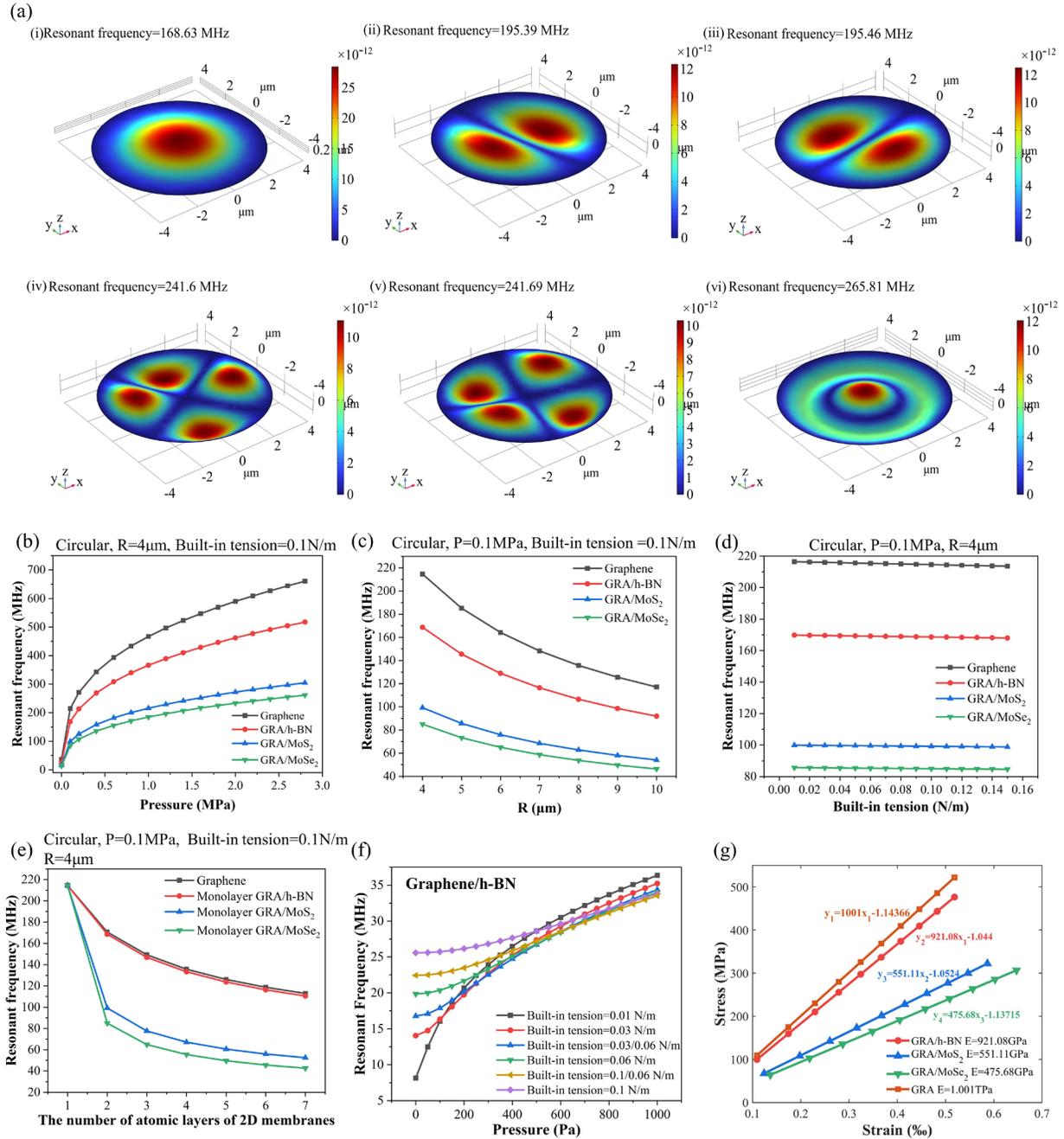

**Figure 5**. Resonant frequencies and Young's moduli of circular 2D membranes of graphene and graphene-based heterostructures under different conditions. (a)The first (i), second (ii), third (iii), fourth(iv), fifth(v), and sixth(vi) order resonant modes of the heterostructures of monolayer graphene with monolayer h-BN. The corresponding resonant frequencies were 168.63 MHz, 195.39 MHz, 195.46 MHz, 241.6 MHz, 241.69 MHz and 265.81 MHz, respectively. (b) The resonant frequencies of graphene and 2D heterostructures of monolayer graphene with monolayer h-BN, MoS$_2$ and MoSe$_2$ versus applied pressures under the conditions of the built-in tension of 0.1 N/m and the 2D membrane's radius of 4 μm. (c) The resonant frequencies of monolayer graphene and 2D heterostructures of monolayer graphene with monolayer h-BN, MoS$_2$ and MoSe$_2$ versus the radiuses of 2D membranes under the conditions of the applied pressure of 0.1 MPa and the built-in tension of 0.1 N/m. (d) The resonant frequencies of monolayer graphene and 2D heterostructures of monolayer graphene with monolayer h-BN, MoS$_2$ and MoSe$_2$ versus built-in tensions under the conditions of the applied pressure of 0.1 MPa and the 2D membranes' radius of 4 μm. (e) The resonant frequencies of graphene membranes with different atomic layers and 2D heterostructures of monolayer graphene with different atomic-layers of h-BN, MoS$_2$ and MoSe$_2$ under the conditions of the applied pressure of 0.1 MPa, built-in tension of 0.1 N/m, and the 2D membranes' radius of 4 μm. It should be noted that as the number of atomic layers of 2D membranes is 1, only monolayer graphene membrane is present without heterostructures. (f) The resonant frequencies of the heterostructure of monolayer graphene with monolayer h-BN versus the applied pressures at different built-in tensions. (g) The data points of stress versus strain of monolayer graphene and graphene-based heterostructures. The curve fittings show the Young's moduli of the monolayer graphene and the heterostructures of monolayer graphene with monolayer h-BN, MoS$_2$ and MoSe$_2$ are 1.001TPa, 921.08 GPa, 551.11 GPa, and 475.68 GPa, respectively.

**Young's Moduli of 2D Membranes**

We simulated the in-plane components of stresses and strains under the conditions of a series of applied forces, and the data points of the strain-stress were plotted and fitted (**Figure** 5g). According to Young's modulus, $E = \sigma/\varepsilon$, the slopes of the fitted curves in **Figure 5**g are values of Young's moduli of graphene and graphene-based heterostructures. It should be noted that there is still a strain in the 2D membranes as the externally applied in-plane stress is zero due to the non-ignorable built-in stress of such 2D membranes. The Young's moduli of the graphene and the heterostructures of monolayer graphene with monolayer h-BN, $MoS_2$, and $MoSe_2$ are estimated to be 1.001TPa, 921.08 GPa, 551.11 GPa, and 475.68 GPa, respectively, which are similar to the values that were experimentally measured in the previous studies (**Table 2**).[38–40] This further verifies the accuracy of our simulation in terms of estimation of Young's modulus of 2D heterostructures. It should be noted that Young's moduli of the heterostructures of monolayer graphene with monolayer h-BN, $MoS_2$, and $MoSe_2$ have been minimally reported. Further, according to our investigations, Young's modulus of graphene/$MoSe_2$ has not been reported before. Therefore, our simulation results reveal Young's moduli of graphene-based heterostructures.

**Table 2**. Comparison of Young's moduli of monolayer graphene and graphene-based heterostructures obtained by our simulation method and previously reported experimental methods.

| 2D materials | Our simulation (GPa) | Experiment (GPa) |
| --- | --- | --- |
| Graphene | 1001 | 1000[2] <br> 1000[38] |
| Graphene/h-BN | 921.08 | 946.97[39] |
| Graphene/$MoS_2$ | 551.11 | 475.13[39] <br> 530[40] |
| Graphene/$MoSe_2$ | 475.68 | - |

**Asymmetric cavities**

In the actual manufacturing process, the asymmetric cavities are often not avoided. The purpose of this section is to simulate the mechanical properties of two typical asymmetric shapes of cavities including elliptical cavity and hexagonal cavity under different loading conditions. We simulated the maximum deflection, maximum strain and resonant frequency of elliptical membrane, hexagonal membrane, and circular membrane of monolayer graphene/ monolayer h-BN heterostructures that have the similar areas under the conditions of a series of applied pressure (**Figure 6**). The area of the elliptical membrane is about 20% larger than the circular membrane while the hexagonal membrane is about 20% smaller than the circular membrane. As a result, the elliptical membrane of monolayer graphene/ monolayer h-BN heterostructures shows the highest value of the maximum deflection among three types of membranes under the condition of the same applied pressure (**Figure 6**a). In contrast, the hexagonal membrane of monolayer graphene/ monolayer h-BN heterostructures has the lowest value of the maximum deflection among three types of membranes under the condition of the same applied pressure (**Figure 6**a). Likewise, among three types of membranes under the condition of the same applied pressure, the value of the maximum strain in the elliptical membrane of monolayer graphene/monolayer h-BN heterostructures is the largest, while the value of the maximum strain in the hexagonal membrane of graphene/h-BN heterostructures is lowest (**Figure 6**b). In terms of resonant frequency, the hexagonal membrane of graphene/h-BN heterostructures has the highest resonant frequency while the elliptical membrane of graphene/h-BN heterostructures has the lowest resonant frequency (**Figure 6**c). Figure 6 d and e show the deflection modeling of elliptical and hexagonal 2D membranes of suspended monolayer graphene/monolayer h-BN heterostructures, respectively.

The first, second, third, and fifth order resonant modes of the ellipsoidal membrane of the heterostructures of monolayer graphene with monolayer h-BN are similar to those of the circular membrane. But the fourth and sixth order resonant mode of the ellipsoidal membrane of the heterostructures of monolayer graphene with monolayer h-BN show a complex vibration pattern with a resonant frequency of 218.7 MHz and 251.9 MHz, compared to the circular membrane (**Figure 6**f, g). To be specific, the sixth order resonant mode exhibits a distribution of multiple vibrational nodes and antinodes, and is more pronounced in the direction of the long axis. This complex resonant mode indicates the non-ignorable impact of the geometrical asymmetry of the elliptical membrane on the vibrational resonant modes, with different stress distributions in the long-axis and short-axis directions (**Figure 6**f, g).

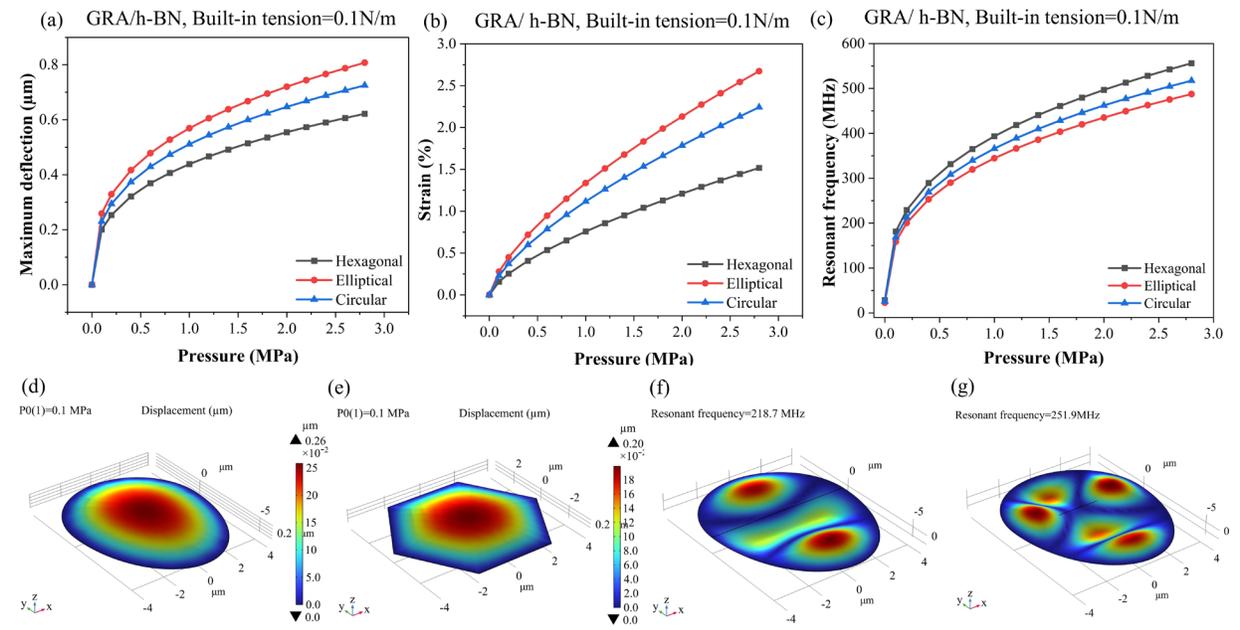

**Figure 6.** The maximum deflection (a), maximum strain (b), and resonant frequency (c) of different shapes of 2D membranes of monolayer graphene/ monolayer h-BN heterostructures

versus applied pressures under the conditions of the built-in tension of 0.1 N/m. Deflection modeling of 2D membranes of suspended monolayer graphene/monolayer h-BN heterostructures with the different membrane's shapes: (d) elliptical membrane (e) hexagonal membrane. The fourth (f) and sixth (g) order resonant mode of the elliptical membrane of the monolayer graphene with monolayer h-BN heterostructures.

**CONCLUSION**

In conclusion, we studied the factors that impact the mechanical properties of suspended graphene-based heterostructures that could be potentially used for ultra-small and high-performance piezoresistive and resonant NEMS pressure sensors. We used COMSOL finite element simulations to simulate and analyze the deflections, strains, resonant frequencies and Young's moduli of the graphene-based heterostructures. For the same surface area, the circular membranes of graphene-based heterostructures have the largest responsivity to the applied pressures compared to the rectangular and square membranes. Both the deflections and strains of suspended graphene-based heterostructures increased with increasing the applied pressures and geometrical sizes, but decreased with increasing the built-in tensions and the number of atomic layers. The resonant frequencies of suspended graphene-based heterostructures increased with increasing the applied pressures, but decreased with increasing geometrical sizes and the number of atomic layers. More important, we estimated Young's moduli of graphene and graphene-based heterostructures, including 1.001TPa, 921.08 GPa, 551.11 GPa, and 475.68 GPa for graphene, graphene/h-BN, graphene/$MoS_2$, and graphene/$MoSe_2$, respectively. Finally, we discuss the highly asymmetric of the cavity structure of the two-dimensional material exhibited during fabrication, where the deformation patterns and mechanical properties under applied pressure are significantly different from those of a symmetric circular membrane. These findings would help to understand the mechanical properties of suspended 2D membranes of graphene-based heterostructures and provide theoretical support and guidance for the design, manufacture and application of next-generation 2D NEMS pressure sensors.

## AUTHOR INFORMATION

### Corresponding Authors

*E-mail: xgfan@bit.edu.cn, jie.ding@bit.edu.cn, wdzhang@nuc.edu.cn

### Author Contributions

Conceptualization, investigation and simulation, writing—original draft preparation, Q.L.; simulation, C.H; conceptualization, supervision, writing—review and editing, J.D., W.Z and X.F.

### Notes

Xuge Fan and Jie Ding are co-inventors on two patents' application about NEMS pressure sensors based on graphene-based heterostructures (Application numbers: 202310222727.3; 202311729735.3). The other authors have no competing financial interest.

## ACKNOWLEDGMENT

This work was supported by the National Natural Science Foundation of China (Grant No. 62171037 and 62088101), 173 Technical Field Fund (2023-JCJQ-JJ-0971), Beijing Natural Science Foundation (4232076), National Key Research and Development Program of China (2022YFB3204600), Beijing Institute of Technology Science and Technology Innovation Plan, National Science Fund for Excellent Young Scholars (Overseas), Beijing Institute of Technology Teli Young Fellow Program (2021TLQT012).

# REFERENCES

(1) Ni, Z. H.; Wang, H. M.; Kasim, J.; Fan, H. M.; Yu, T.; Wu, Y. H.; Feng, Y. P.; Shen, Z. X. Graphene Thickness Determination Using Reflection and Contrast Spectroscopy. *Nano Lett.* **2007**, *7* (9), 2758–2763. https://doi.org/10.1021/nl071254m.

(2) Lee, C.; Wei, X.; Kysar, J. W.; Hone, J. Measurement of the Elastic Properties and Intrinsic Strength of Monolayer Graphene. *Science* **2008**, *321* (5887), 385–388. https://doi.org/10.1126/science.1157996.

(3) Cocco, G.; Cadelano, E.; Colombo, L. Gap Opening in Graphene by Shear Strain. *Phys. Rev. B* **2010**, *81* (24), 241412. https://doi.org/10.1103/PhysRevB.81.241412.

(4) Bolotin, K. I.; Sikes, K. J.; Jiang, Z.; Klima, M.; Fudenberg, G.; Hone, J.; Kim, P.; Stormer, H. L. Ultrahigh Electron Mobility in Suspended Graphene. *Solid State Communications* **2008**, *146* (9), 351–355. https://doi.org/10.1016/j.ssc.2008.02.024.

(5) Koenig, S. P.; Boddeti, N. G.; Dunn, M. L.; Bunch, J. S. Ultrastrong Adhesion of Graphene Membranes. *Nature Nanotech* **2011**, *6* (9), 543–546. https://doi.org/10.1038/nnano.2011.123.

(6) Lemme, M. C.; Wagner, S.; Lee, K.; Fan, X.; Verbiest, G. J.; Wittmann, S.; Lukas, S.; Dolleman, R. J.; Niklaus, F.; Van Der Zant, H. S. J.; Duesberg, G. S.; Steeneken, P. G. Nanoelectromechanical Sensors Based on Suspended 2D Materials. *Research* **2020**, *2020*, 2020/8748602. https://doi.org/10.34133/2020/8748602.

(7) Zhang, Z.; Liu, Q.; Ma, H.; Ke, N.; Ding, J.; Zhang, W.; Fan, X. Recent Advances in Graphene-Based Pressure Sensors: A Review. *IEEE Sensors Journal* **2024**, 1–1. https://doi.org/10.1109/JSEN.2024.3419243.

(8) Smith, A. D.; Niklaus, F.; Paussa, A.; Vaziri, S.; Fischer, A. C.; Sterner, M.; Forsberg, F.; Delin, A.; Esseni, D.; Palestri, P.; Östling, M.; Lemme, M. C. Electromechanical Piezoresistive Sensing in Suspended Graphene Membranes. *Nano Lett.* **2013**, *13* (7), 3237–3242. https://doi.org/10.1021/nl401352k.

(9) Smith, A. D.; Vaziri, S.; Delin, A.; Ostling, M.; Lemme, M. C. Strain Engineering in Suspended Graphene Devices for Pressure Sensor Applications. In *2012 13th International Conference on Ultimate Integration on Silicon (ULIS)*; IEEE: Grenoble, France, 2012; pp 21–24. https://doi.org/10.1109/ULIS.2012.6193347.

(10) Dolleman, R. J.; Davidovikj, D.; Cartamil-Bueno, S. J.; van der Zant, H. S. J.; Steeneken, P.

G. Graphene Squeeze-Film Pressure Sensors. *Nano Lett* **2016**, *16* (1), 568–571. https://doi.org/10.1021/acs.nanolett.5b04251.

(11) Chen, Y.-M.; He, S.-M.; Huang, C.-H.; Huang, C.-C.; Shih, W.-P.; Chu, C.-L.; Kong, J.; Li, J.; Su, C.-Y. Ultra-Large Suspended Graphene as a Highly Elastic Membrane for Capacitive Pressure Sensors. *Nanoscale* **2016**, *8* (6), 3555–3564. https://doi.org/10.1039/C5NR08668J.

(12) Davidovikj, D.; Scheepers, P. H.; Van Der Zant, H. S. J.; Steeneken, P. G. Static Capacitive Pressure Sensing Using a Single Graphene Drum. *ACS Appl. Mater. Interfaces* **2017**, *9* (49), 43205–43210. https://doi.org/10.1021/acsami.7b17487.

(13) Lee, M.; Davidovikj, D.; Sajadi, B.; Šiškins, M.; Alijani, F.; Van Der Zant, H. S. J.; Steeneken, P. G. Sealing Graphene Nanodrums. *Nano Lett.* **2019**, *19* (8), 5313–5318. https://doi.org/10.1021/acs.nanolett.9b01770.

(14) Šiškins, M.; Lee, M.; Wehenkel, D.; Van Rijn, R.; De Jong, T. W.; Renshof, J. R.; Hopman, B. C.; Peters, W. S. J. M.; Davidovikj, D.; Van Der Zant, H. S. J.; Steeneken, P. G. Sensitive Capacitive Pressure Sensors Based on Graphene Membrane Arrays. *Microsyst Nanoeng* **2020**, *6* (1), 102. https://doi.org/10.1038/s41378-020-00212-3.

(15) Romijn, J.; Dolleman, R. J.; Singh, M.; Zant, H. S. J. van der; Steeneken, P. G.; Sarro, P. M.; Vollebregt, S. Multi-Layer Graphene Pirani Pressure Sensors. *Nanotechnology* **2021**, *32* (33), 335501. https://doi.org/10.1088/1361-6528/abff8e.

(16) Zhu, S.-E.; Ghatkesar, M.; Zhang, C.; Janssen, G. Graphene Based Piezoresistive Pressure Sensor. *Applied Physics Letters* **2013**, *102*. https://doi.org/10.1063/1.4802799.

(17) Wang, Q.; Hong, W.; Dong, L. Graphene "Microdrums" on a Freestanding Perforated Thin Membrane for High Sensitivity MEMS Pressure Sensors. *Nanoscale* **2016**, *8* (14), 7663–7671. https://doi.org/10.1039/C5NR09274D.

(18) Berger, C.; Phillips, R.; Centeno, A.; Zurutuza, A.; Vijayaraghavan, A. Capacitive Pressure Sensing with Suspended Graphene–Polymer Heterostructure Membranes. *Nanoscale* **2017**, *9* (44), 17439–17449. https://doi.org/10.1039/C7NR04621A.

(19) Lin, X.; Liu, Y.; Zhang, Y.; Yang, P.; Cheng, X.; Qiu, J.; Liu, G. Polymer-Assisted Pressure Sensor with Piezoresistive Suspended Graphene and Its Temperature Characteristics. *NANO* **2019**, *14* (10), 1950130. https://doi.org/10.1142/S1793292019501303.

(20) Xu, J.; Wood, Graham. S.; Mastropaolo, E.; Newton, Michael. J.; Cheung, R. Realization of a Graphene/PMMA Acoustic Capacitive Sensor Released by Silicon Dioxide Sacrificial

Layer. *ACS Appl. Mater. Interfaces* **2021**, *13* (32), 38792–38798. https://doi.org/10.1021/acsami.1c05424.

(21) Androulidakis, C.; Zhang, K.; Robertson, M.; Tawfick, S. Tailoring the Mechanical Properties of 2D Materials and Heterostructures. *2D Mater.* **2018**, *5* (3), 032005. https://doi.org/10.1088/2053-1583/aac764.

(22) Manzeli, S.; Allain, A.; Ghadimi, A.; Kis, A. Piezoresistivity and Strain-Induced Band Gap Tuning in Atomically Thin MoS2. *Nano Lett* **2015**, *15* (8), 5330–5335. https://doi.org/10.1021/acs.nanolett.5b01689.

(23) Hosseini, M.; Elahi, M.; Pourfath, M.; Esseni, D. Very Large Strain Gauges Based on Single Layer MoSe2 and WSe2 for Sensing Applications. *Applied Physics Letters* **2015**, *107* (25), 253503. https://doi.org/10.1063/1.4937438.

(24) Huang, M.; Pascal, T. A.; Kim, H.; Goddard, W. A. I.; Greer, J. R. Electronic−Mechanical Coupling in Graphene from in Situ Nanoindentation Experiments and Multiscale Atomistic Simulations. *Nano Lett.* **2011**, *11* (3), 1241–1246. https://doi.org/10.1021/nl104227t.

(25) Wagner, S.; Yim, C.; McEvoy, N.; Kataria, S.; Yokaribas, V.; Kuc, A.; Pindl, S.; Fritzen, C.-P.; Heine, T.; Duesberg, G. S.; Lemme, M. C. Highly Sensitive Electromechanical Piezoresistive Pressure Sensors Based on Large-Area Layered PtSe2 Films. *Nano Lett* **2018**, *18* (6), 3738–3745. https://doi.org/10.1021/acs.nanolett.8b00928.

(26) Morell, N.; Reserbat-Plantey, A.; Tsioutsios, I.; Schädler, K. G.; Dubin, F.; Koppens, F. H. L.; Bachtold, A. High Quality Factor Mechanical Resonators Based on WSe2 Monolayers. *Nano Lett.* **2016**, *16* (8), 5102–5108. https://doi.org/10.1021/acs.nanolett.6b02038.

(27) Will, M.; Hamer, M.; Müller, M.; Noury, A.; Weber, P.; Bachtold, A.; Gorbachev, R. V.; Stampfer, C.; Güttinger, J. High Quality Factor Graphene-Based Two-Dimensional Heterostructure Mechanical Resonator. *Nano Lett.* **2017**, *17* (10), 5950–5955. https://doi.org/10.1021/acs.nanolett.7b01845.

(28) Lee, J.; Wang, Z.; He, K.; Yang, R.; Shan, J.; Feng, P. X.-L. Electrically Tunable Single- and Few-Layer MoS2 Nanoelectromechanical Systems with Broad Dynamic Range. *Science Advances* **2018**, *4* (3), eaao6653. https://doi.org/10.1126/sciadv.aao6653.

(29) Xu, Y.; Guo, Z.; Chen, H.; Yuan, Y.; Lou, J.; Lin, X.; Gao, H.; Chen, H.; Yu, B. In-Plane and Tunneling Pressure Sensors Based on Graphene/Hexagonal Boron Nitride Heterostructures. *Applied Physics Letters* **2011**, *99* (13), 133109. https://doi.org/10.1063/1.3643899.

(30) Kim, S. J.; Mondal, S.; Min, B. K.; Choi, C.-G. Highly Sensitive and Flexible Strain–Pressure Sensors with Cracked Paddy-Shaped MoS2/Graphene Foam/Ecoflex Hybrid Nanostructures. *ACS Appl. Mater. Interfaces* **2018**, *10* (42), 36377–36384. https://doi.org/10.1021/acsami.8b11233.

(31) Ruiz-Vargas, C. S.; Zhuang, H. L.; Huang, P. Y.; van der Zande, A. M.; Garg, S.; McEuen, P. L.; Muller, D. A.; Hennig, R. G.; Park, J. Softened Elastic Response and Unzipping in Chemical Vapor Deposition Graphene Membranes. *Nano Lett* **2011**, *11* (6), 2259–2263. https://doi.org/10.1021/nl200429f.

(32) Fan, X.; Forsberg, F.; Smith, A. D.; Schröder, S.; Wagner, S.; Rödjegård, H.; Fischer, A. C.; Östling, M.; Lemme, M. C.; Niklaus, F. Graphene Ribbons with Suspended Masses as Transducers in Ultra-Small Nanoelectromechanical Accelerometers. *Nat Electron* **2019**, *2* (9), 394–404. https://doi.org/10.1038/s41928-019-0287-1.

(33) Fan, X.; Niklaus, F. Deformation Behavior and Mechanical Properties of Suspended Double-Layer Graphene Ribbons Induced by Large Atomic Force Microscopy Indentation Forces. *Advanced Engineering Materials* **2022**, *24* (3), 2100826. https://doi.org/10.1002/adem.202100826.

(34) Fan, X.; Moreno-Garcia, D.; Ding, J.; Gylfason, K. B.; Villanueva, L. G.; Niklaus, F. Resonant Transducers Consisting of Graphene Ribbons with Attached Proof Masses for NEMS Sensors. *ACS Appl. Nano Mater.* **2024**, *7* (1), 102–109. https://doi.org/10.1021/acsanm.3c03642.

(35) Moreno-Garcia, D.; Fan, X.; Smith, A. D.; Lemme, M. C.; Messina, V.; Martin-Olmos, C.; Niklaus, F.; Villanueva, L. G. A Resonant Graphene NEMS Vibrometer. *Small* **2022**, *18* (28), 2201816. https://doi.org/10.1002/smll.202201816.

(36) Tabata, O.; Kawahata, K.; Sugiyama, S.; Igarashi, I. Mechanical Property Measurements of Thin Films Using Load-Deflection of Composite Rectangular Membrane. In *IEEE Micro Electro Mechanical Systems, , Proceedings, "An Investigation of Micro Structures, Sensors, Actuators, Machines and Robots";* IEEE: Salt Lake City, UT, USA, 1989; pp 152–156. https://doi.org/10.1109/MEMSYS.1989.77981.

(37) Bunch, J. S.; Verbridge, S. S.; Alden, J. S.; van der Zande, A. M.; Parpia, J. M.; Craighead, H. G.; McEuen, P. L. Impermeable Atomic Membranes from Graphene Sheets. *Nano Lett.* **2008**, *8* (8), 2458–2462. https://doi.org/10.1021/nl801457b.

(38) Tan, X.; Wu, J.; Zhang, K.; Peng, X.; Sun, L.; Zhong, J. Nanoindentation Models and Young's Modulus of Monolayer Graphene: A Molecular Dynamics Study. *Applied Physics Letters* **2013**, *102* (7), 071908. https://doi.org/10.1063/1.4793191.

(39) Li, X.; Huang, C.; Hu, S.; Deng, B.; Chen, Z.; Han, W.; Chen, L. Negative and Near-Zero Poisson's Ratios in 2D Graphene/$MoS_2$ and Graphene/h-BN Heterostructures. *J. Mater. Chem. C* **2020**, *8* (12), 4021–4029. https://doi.org/10.1039/C9TC06424A.

(40) Elder, R. M.; Neupane, M. R.; Chantawansri, T. L. Stacking Order Dependent Mechanical Properties of Graphene/MoS2 Bilayer and Trilayer Heterostructures. *Applied Physics Letters* **2015**, *107* (7), 073101. https://doi.org/10.1063/1.4928752.